\documentclass[pra,tightenlines,a4paper,preprint,amsmath,aps]{revtex4}
\usepackage{dcolumn,bm}

\begin{document}

\title{\small NEUTRON DIFFRACTION,MAGNETIZATION AND ESR STUDIES
OF PSEUDOCUBIC Nd$_{0.75}$Ba$_{0.25}$MnO$_{3}$ AND ITS UNUSUAL
CRITICAL BEHAVIOR ABOVE $T_C$}
\author{V.A.~Ryzhov, A.V.~Lazuta, O.P.~Smirnov, I.A.~Kiselev,
Yu.P.~Chernenkov, and S.A.~Borisov}
\address{Petersburg Nuclear Physics Institute, Gatchina,
Leningrad district, 188300, Russia}
\author{I.O.~Troaynchuk and D.D.~Khalyavin}
\address{Institute of Physics of Solids and Semiconductors,
National Academy of Sciences, ul. P. Brovki 17, Minsk, 220072,
Belarus}
\date{recieved; accepted}

\begin{abstract}
Results of structural neutron diffraction study, magnetization
and electron spin resonance (ESR) measurements are presented for
insulating Nd$_{1-x}$Ba$_{x}$MnO$_3$ ($x$ = 0.25) with the Curie
temperature $T_C \approx$ 129 K. Its pseudocubic structure
reveals the definite distortions to a lower symmetry. Detailed
analysis of the data is performed by using $Pbnm$ space group in
a temperature range 4.2 $\div$ 300 K. The compound is found to
exhibit the Jahn-Teller (JT) transition at $T_{JT} \approx$ 250
K. Character of the coherent JT distortions and their temperature
evolution differ from those of the $x$ = 0.23 manganite. The
field cooled magnetization data are in a reasonable agreement
with the predictions for a 3D isotropic ferromagnet above $T_C$.
These measurements, however, reveal a difference between the
field cooled and zero field cooled data in the paramagnetic
region. The ESR results are also in a correspondence with
behavior of a 3D isotropic ferromagnet above $T^{\ast} \approx$
143 K ($\tau^{\ast} \approx 0.12 \leq \tau < 1, \tau$ =
$(T-T_C)/T_C$). It is shown that an anisotropic exchange coupling
of the Mn and Nd magnetic moments can give a substantial
contribution in ESR linewidth masking its critical enhancement.
The different temperature treatments (slow/fast cooling/heating,
with/without external magnetic field) of the sample reveal a
temperature hysteresis of the ESR spectra below $T^{\ast}$
indicating an anomalous response in the paramagnetic region. The
study of the magnetic phase transition in the $x$ = 0.23 and 0.25
manganites suggests change in its character from the second to
first order at $T^{\ast}$. The conventional free energy including
the magnetization and magnetic field is not found to describe
this first order transition. This suggests that the charge,
orbital and JT phonon degrees of freedom, in addition to
magnetization, may be the critical variables, the unusual
character of the transition being determined by their coupling.
The unconventional critical behavior is attributed to an orbital
liquid metallic phase that begins to coexist with the initial
orbital ordered phase below $T^{\ast}$.
\end{abstract}

\maketitle

%Keywords: A: magnetically ordered materials, D: phase
%transitions, E: neutron scattering, E: electron paramagnetic
%resonance
%PACS: 61.12-q, 75.40.Gb, 76.30.

\newpage

%\begin{twocolumn}{2}
%\narrowtext
%\twocolumn

\section {\small Introduction}

Doped manganites L$_{1-x}$A$_{x}$MnO$_3$ (L and A are rare earth
and alkaline earth ions) usually exhibits the ferromagnetic (F)
ground state in a region $x_0 \leq x \leq$ 0.5. In
Nd$_{1-x}$Ba$_x$MnO$_3$, this state develops below $T_C \sim$ 120
K for 0.2 $\leq x \leq$ 0.35 \cite{TKS}. For $x <$ 0.3, these
compounds show insulating (I) behavior in the ordered state,
whereas for $x \geq$ 0.3 the F metallic (M) state is observed
below $T_C$. A colossal magnetoresistance occurs for all $x$
near $T_C$. The unexpected coexistence of ferromagnetic and
insulating behavior cannot be explained by the double exchange
mechanism and remains to be poor understood. A possible reason
of the charge localization can be a long-range orbital ordering
due to Jahn-Teller (JT) effect. This source of the localization
was suggested to account for the FI ground state in
La$_{1-x}$Ca$_x$MnO$_3$ for 0.12 $< x <$ 0.2 \cite{VJM}. The FI
state can also be consequence of a spontaneous orbital ordering,
which develops in a cubic manganite, due to a coupling of the
spin and orbital degrees of freedom arising from the electron
correlations \cite{OIM}. La$_{0.88}$Sr$_{0.12}$MnO$_3$ was found
to exhibit the FI state of such nature \cite{EHI}. Note that a
new explanation of the insulator to metal transition, which
occurs in an orbital disordered state, has been suggested
recently \cite{RKH}. This approach permits the existence an
orbital liquid insulating state. As far as we know, however,
such a state has not been found so far.

An important aspect of this problem is a nature of the
paramagnetic (P) -- FI transition which can differ from that of a
3D isotropic ferromagnet. A few works have only concentrated on
behavior at this critical point. An unconventional P-FI
transition was observed in insulating La$_{0.8}$Ca$_{0.2}$MnO$_3$
where a magnetic correlation length increased only from 10 to 35
\AA \; as $T \rightarrow T_C$ \cite{DFP}. To investigate this
issue in detail, we have used Nd$_{1-x}$Ba$_x$MnO$_3$ compounds,
which are closed to the cubic perovskites due to the small
structural distortions \cite{TKS}. Also of interest in this
system is the effect of a Mn-Nd interaction on its properties.
In Nd$_{0.7}$Ba$_{0.3}$MnO$_3$, the ordering of the Nd moments,
that develops below 20 K \cite{FSM}, changed the magnetic and
transport properties \cite{MMH}. As study of the
Nd$_{1-x}$Sr$_x$MnO$_3$ (0.1 $\leq x \leq$ 0.5) showed, the Nd
magnetic subsystem came also into play at the low temperatures
due to its ordering \cite{PMR}. At the same time, no effects of
the Mn-Nd coupling were found near $T_C$ in the investigations
of critical static behavior of Nd$_{0.6}$Pb$_{0.4}$MnO$_3$
\cite{SRN} and critical dynamics of Nd$_{1-x}$Sr$_x$MnO$_3$ ($x
\approx$ 0.5) \cite{KWN}. Our studies have included the
measurements of the linear and nonlinear susceptibilities as
well as ESR. The neutron diffraction investigations have been
also performed to elucidate the effects of the lattice
distortions and orbital ordering. The study of the critical
magnetic properties of the $x$ = 0.23 ($T_C \approx$ 124 K) and
$x$ = 0.25 ($T_C \approx$ 129 K) manganites revealed a more
complicated scenario than that expected for the 3D isotropic
ferromagnets \cite{LKR,RLL,RLK,RyL}. The conventional critical
behavior was observed to proceed for $T^{\ast} < T < 2T_C$ where
$T^{\ast} \approx$ 147 K ($x$ = 0.23) and $T^{\ast} \approx$ 143
K ($x$ = 0.25). In contrast, the data on nonlinear response of
the second order clearly indicated the occurrence of an
anomalous critical behavior for $T_C < T < T^{\ast}$ that
suggested the coexistence of the two magnetic phases. Although
the $x$ = 0.23 compound exhibited an orbital ordering due to the
JT effect below $T_{JT} \sim$ 350K, the neutron diffraction
study did not reveal a two phase structure in the anomalous
$T-$range \cite{RLK} that would be a natural explanation of the
phase separated (PS) magnetic state.  In addition, according to
the ESR measurements, the transition did not become hysteretic
in temperature and clearly first order.

In this paper we present the neutron diffraction, magnetization
and ESR investigations of $x$ = 0.25 manganite. The $T_{JT}$ is
reduced on doping, becoming $T_{JT} \approx$ 250 K. A character
of the JT distortions and orbital ordering differs from that of
the $x$ = 0.23 compound. $T$-evolution of the structural
parameters does not show any anomalies that can be definitely
attributed to formation of a structural PS state in the anomalous
$T$-range. The magnetization, $M(H)$, measurements are performed
in the field cooled (FC) and zero field cooled (ZFC) regimes at
$H =$ 1 kOe in a range 4.2 $\div$ 300 K. The FC data agree
reasonably with prediction for a 3D isotropic ferromagnet in the
paramagnetic critical region from 2$T_C$ down to $T_C$. At the
same time, a difference between the FC and ZFC results is
unexpectedly found to proceed up to 300 K. ESR measurements are
carried out above $T_C$. A relaxation rate of uniform
magnetization $\Gamma$ and $g$-factor show nearly the same
dependencies above $T^{\ast}$ as for the $x =$ 0.23 manganite.
The $\Gamma(\tau)$ does not exhibit a critical enhancement as it
occurs in the conventional cubic ferromagnets CdCr$_2$S$_4$ and
CdCr$_2$Se$_4$ \cite{BIL}, as well as in the near to cubic
manganites La$_{1-x}$Ca$_x$MnO$_3$ ($x = 0.18 \div 0.22$
\cite{SRG}.  An anisotropic exchange Mn-Nd coupling is shown can
give a substantial contribution in $\Gamma$, masking the
critical enhancement. In order to examine character of the phase
transition, effect of the different temperature treatments of the
sample on the ESR signals is elucidated. These measurements and
$M(H)$ data support the existence of the magnetic PS state near
$T_C$. The possible reasons of origination of the heterophase
magnetic state are discussed.

\section {Samples and methods}

The polycrystalline samples were prepared by the standard
technology \cite{TKS}, an oxygen content being controlled by a
thermogravimetric analysis. A 48-counter powder neutron
diffractometer of PNPI with a wave length 1.83 \AA \; of an
incident neutron beam was used in the structural investigations.
The measurements were performed in a cryostat at 4.2 $\div$ 300
K.  Program FULLPROF was employed for the structure refinement.

The magnetization measurements were carried out using a SQUID
magnetometer in a temperature range 4.2 $\div$ 300 K and a
magnetic field 1 kOe. A single crystal with a mass {\$}m{\$} =
91.8 mg was used that came from the same growth batch as the
crystal of work \cite{RyL}.

In ESR study, the small part of single crystal investigated in
work \cite{RyL} with a mass $m \approx$ 1.6 mg was used. We
employed a special $X$-band ESR spectrometer ($f =$ 8.37 GHz)
described earlier \cite{RZS,RLa}. It registers a component of
magnetization of a sample which is proportional to an
off-diagonal part of its magnetic susceptibility ($M_y(\omega) =
\chi_{yx}(\omega)h_x(\omega)$) when a linear polarized exciting
$ac-$field {\bf h}$\|$ Ox and a steady field {\bf H}$\|$ Oz.

Below $T^{\ast} \approx$ 143 K the nonlinear response of the
compound reveals features of the PS magnetic state that can
signal the first order transition \cite{RyL}. It is natural,
therefore, to expect a temperature hysteresis of the ESR signal.
Besides, a state forming at the PS usually depends on a process
that is used to transfer a system in the PS regime. Therefore,
the several types of the temperature scan without/with an
external magnetic field were performed.

\begin{enumerate}

\item The sample was slowly cooled from room temperature down to
128 K below $T_C$ and then slowly heated back to room
temperature. In the chosen temperature points ($T-$points) the
ESR spectra were registered. A temporal stabilization time before
recording a signal was about 200-300 seconds at each $T-$points.

\item The same temperature scan as above was applied but with
$H$ = 4 kOe during cooling between $T-$points. This procedure
also tests a possible PS state because the magnetic field can
change a balance between the phase fractions.

\item A fast cooling of the sample was used down to 128 K in
zero $H$ and then the sample was slow heated up to room
temperature.

\end{enumerate}

\section {Structural data}

A neutron diffraction pattern at room temperature is plotted in
Fig.1. Similar to Nd$_{0.77}$Ba$_{0.23}$MnO$_3$ crystal
\cite{RLK}, pseudocubic structure of
Nd$_{0.75}$Ba$_{0.25}$MnO$_3$ exhibits distortions to a
monoclinic symmetry, as the characteristic peaks at the low
angles of scattering evidence (see the insert in Fig.1). They
unambiguously originate from a lattice deformation because the
X-ray diffraction patterns also show them at room temperature. A
width of the peaks is larger than that of the main Bragg
reflections. It indicates that the regions of the relatively
small sizes possess the monoclinic distortions. The peaks
exhibit no temperature variations and have a small amplitude.
The presence of small amount of such a monoclinic phase is
expected to have a minimal effect on the $T-$dependent
properties of this compound. The measured diffraction profiles
are found to be well fitted using the orthorhombic $Pbnm$ space
group. This fit gives only the slightly worse description than
that of the monoclinic group $P21/m$. Therefore, the further
analysis will be based on the $Pbnm$ setting. Fig.2 displays the
$T-$variations of structural parameters. The orthorhombic
distortions are small in range of the measurements as in the $x
=$ 0.23 manganite \cite{RLK}. However, there are essential
differences in their character.

The data on the Mn-O bond lengths (Fig.2c) indicate that the JT
transition from the high temperature orbital disordered $O$ phase
to the $O^{\prime}$ phase possessing the cooperative JT
distortions occurs at $T_{JT} \sim$ 250 K. This transition
involves the strong abrupt changes of the Mn-O-Mn angles between
220 $\div$ 290 K when the curves of monotonic $T-$variation for
the Mn-O1-Mn and Mn-O2-Mn angles cross one another (Fig.2d). This
transformation is related to rotation of the MnO$_6$ octahedra
around the pceudocubic [111] direction \cite{COM}. Thus, the
$OO^{\prime}$ transition is driven by both the JT distortions
that underlie an orbital ordering and a steric effect (the oxygen
ion displacements connected to the octahedrons tilting). The
latter leads to peculiarity of the transition when a relationship
between the lattice parameters ($a > b \approx c/\sqrt{2}$,
Fig.2a) conserves above and below $T_{JT}$, the splitting $a-b$
becoming even smaller on cooling below $T_{JT}$. The
$OO^{\prime}$ transition is not usually accompanied by the
octahedral tilting. In this case $c/\sqrt{2} \approx a \approx b$
above $T_{JT}$ and $c/\sqrt{2} < a, b$ below it, as it is found
in LaSr \cite{COM}, LaCa \cite{HSL} and $x =$ 0.23 NdBa
\cite{RLK} manganites. The $T-$variations of the lattice
parameters and unit-cell volume (Fig.2a,b) are observed above
$T_C$. They mainly reflect compression of the lattice on cooling
which is related to reduction of the lattice parameter. This
process completes at $T_C$. The Mn-O bound lengths (Fig.2c)
coincide above $T_{JT} \approx$ 250 K and their splitting
remains nearly temperature independent below 230 K, so that the
JT phase forms in the narrow temperature interval. The splitting
corresponds the orbital ordering closed to 3$y^2 - r^2/3x^2 -
r^2$ type which develops in the $a-b$ plane because dMn-O1
$\approx$ dMn-O22 $>$ dMn-O21 \cite{COM}. In
La$_{1-x}$Ca$_x$MnO$_3$ (0.12 $< x \leq$ 0.21) with the FI
ground state, the JT-distortions monotonically increases on
cooling over a large temperature interval \cite{VJM}. At $x =$
0.2, this behavior is observed from $T \approx$ 300 K $> T_C
\approx$ down to 100 K.  It was interpreted as the coexistence
of the $O$ and $O^{\prime}$ phases i.e. as the structural PS
state.  Although our measurements of the dMn-O are not a very
high accuracy, they contradict the coexistence of the $O$ and
$O^{\prime}$ phase with the comparable volume fractions even at
220 K, and support formation of the nearly structural uniform
phase below 230 K.  The $T-$evolution of the structural
parameters does not reveal any peculiarities in the anomalous
$T-$range ($T^{\ast} \approx$ 143 K $\div$ 129 K $\approx T_C$)
which can be directly associated with the appearance of a
two-phase structural state.  Note that a character of the JT
distortions in the $x =$ 0.23 manganite below 150 K (dMn-O21 -
dMn-O1 $\approx$ dMn-O1 - dMnO22), where it depends weakly on
temperature \cite{RLK}, differs from that of the $x =$ 0.25
manganite.  Thus, the FI state and the anomalous critical
behavior can occur at the different types of these distortions.

Temperature dependence of the ferromagnetic moment (Fig.2b) shows
the usual behavior for a ferromagnet with $T_C \approx$ 129 K.
The magnetic moment in the ground state 3.0(1) $\mu$ ($\mu$ is
the Borh magneton) is less than that of the Mn sublattice 3.75
$\mu$ expected for this composition. This is mainly due to
antiparallel alignment of the Nd magnetic moment $\sim 0.5 \mu$
\cite{FSM} that gives the additive contribution in measured
ferromagnetic moment of the sample.

\section{Magnetization}

Fig. 3 displays the temperature dependences of $M$ that are
obtained in the FC and ZFC measurements at $H$ = 1 kOe. We
consider first the FC data above $T_C$. To determine a scaling
behavior of susceptibility of the material $\chi$, it is
convenient to write inverse susceptibility of the sample
$\chi_{ext}^{-1} = H/M$ as
%- - - - - - - - - - - - - - - - - - - - - - - - - - - - - - -
\begin{equation}
\chi_{ext}^{-1} = \chi^{-1} + 4\pi N.
\end{equation}

%- - - - - - - - - - - - - - - - - - - - - - - - - - - - - - -

\noindent
Here $\chi$ =
$C_{\chi}[S(S+1)/(3kT_C)]((g\mu)^2/V_0)\tau^{-\gamma}$ (the
conventional units), $k$ is the Boltzmann constant, $C_{\chi}$ is
the numerical factor, $g$ is the $g-$factor, $V_{0} \approx$
41.45 \AA$^3$ \; is the volume of the unit magnetic cell and $N$ is
the demagnetization factor. A fit of the data for 1 $> \tau >$
0.093 (258 K $> T >$ 141 K) with $T_C \approx $ 129 K and $N$
= 1/3 gives $\gamma =$ 1.39(1), $C_{\chi} =$ 2.95(3) for $S$ =
1.875, these values being independent on $N$ for 0.25 $< N <$ 0.4
(see inset(1) in Fig.3). The same values of $\gamma $ and
$C_{\chi}$ are obtained by employing an interval 1 $> \tau >$
0.256 whereas the fit becomes worse below 141 K. This shows that
$H$ = 1 kOe affects the $\chi(\tau)$ dependence only at $\tau <$
0.093. The $ac$ linear response \cite{RyL} and ESR
measurements (see Fig.5) give a slightly less value of $\gamma
\approx$ 1.32 that is due to the demagnetization effects because
a fit of the $M/H = \chi_{ext}$ data yields also a close
magnitude of $\gamma $.

Since the nonlinear response reveals the anomalous critical
behavior in the low magnetic fields below $T^{\ast} \approx$ 143
K, it is important to check whether the critical $M(\tau)$
dependence at $H$ = 1 kOe is the conventional one or not in the
same $T$-range. Our $M(\tau)$ data in this region are related
mainly to an intermediate $H$-regime between the weak ($T >$ 143
K) and strong ($T = T_C$) magnetic fields. As can be easily
verified, a modified Arrot plot scheme frequently employed for
analysis of the critical $M(\tau,H)$ dependences \cite{DKS,KSR}
has the incorrect analytic properties in the weak fields.
Therefore we use another approximate equation of state
\cite{PaP} with a correct behavior in the weak and strong
fields. It can be presented in a convenient form as \cite{LLR}
%- - - - - - - - - - - - - - - - - - - - - - - - - - - - - - - - - -
\begin{eqnarray}
\frac{M}{H_{int}} \quad = \quad C_0
\Bigl( \frac{\varphi(x)}{\tau} \Bigr)^{\gamma} \Bigl\{ 1-a
\bigl( 1-\varphi(x) \bigr) \Bigr\}^{-1}, \\
x \quad = \quad
h_{int}/\tau^{\beta_H} \quad = \quad A \bigl( 1-\varphi(x)
\bigr)^{1/2} \Bigl\{ 1-a\bigl( 1-\varphi(x)\bigr) \Bigr\}
\varphi(x)^{-\beta_H}.
\end{eqnarray}

%- - - - - - - - - - - - - - - - - - - - - - - - - - - - - - - - - -

\noindent
Here $A >$ 0, 1 $> a >$ 0 are the coefficients, $C_0$ is the
factor determining the amplitude of $\chi$, $\beta_H = \gamma$ +
$\beta$ is the magnetic field index, $\beta$ is the index of the
spontaneous magnetization, $h_{int} = g\mu H_{int}/(kT_C)$ and
$H_{int} = H - 4\pi NM$ is the internal field. The function
$\varphi(x)$ decreases monotonically from 1 down to 0 as $x$
increases from 0 up to $\infty $. The $\varphi(x)$ and,
consequently, $M/H_{int}$ are the even functions of $x$. They can
be expended in the power series in $x^2$ for $x^2 \ll $ 1 because
Eq.(3) can be presented as $x^2 =$ [the right part]$^2$. Strongly
speaking, $a = \gamma/\beta_H$ in this approximation.
Nevertheless, the parameters $A, a$ and $\beta_H$ (or some of
them) can be treated as the free ones at an approximate fit
($\gamma $ has been determined earlier). Insert (2) in Fig.3a
displays the best fit of $M/H_{int}$ at $\gamma \approx$ 1.39,
$\beta_H \approx$ 5/3 (the value for a 3D isotropic ferromagnet)
and $N =$ 1/3 for our sample closed to a cube. It gives $a
\approx$ 0.98 that is not far from the theoretical value $a =
\gamma/\beta_H \approx $ 0.81. We tried also to find
$\beta_H$ using $a = \gamma/\beta_H$. This yields $\beta_H
\approx$ 1.87 ($a \approx$ 0.74), the fit being worse. These
results show that the $M(\tau ,H)$ dependence is comparably well
described by the conventional scaling static theory in the
anomalous region ($T_C \leq T < T^{\ast}$) at $H_{int} \approx H
=$ 1 kOe. Thus, the anomalous nonlinear critical behavior,
that is detected by the $M_2$ measurements in the low magnetic
fields ($H <$ 100 Oe), is suppressed by the stronger field. This
means also that the anomalous phase occupies a relative small
volume (an upper border does not presumably exceed 1/5).

According to the FC and ZFC data, $M(H,T)$ exhibits the
$T-$hysteresis (see Fig.3a and inset in Fig.3b). This phenomenon
is usually observed in the manganaties below $T_C$, whereas its
presence above $T_C$ up to 300 K is the unexpected result. The
$\delta M(T) = 2(M_{\mbox{\small ZFC}} -
M_{\mbox{\small FC}})/(M_{\mbox{\small ZFC}}$ + $M_{\mbox{\small
FC}})$ exhibits the maximum ($T_{m1} \approx$ 290 K, see
Fig.3b) at high temperatures between 220 K and 300 K which can
be attributed to the structural transition occurring in this
$T-$range. Below 220 K, $\delta M$ increases monotonically at $T
\rightarrow T_C+0$.  This means that difference
$M_{\mbox{\small ZFC}}(T)$ - $M_{\mbox{\small FC}}(T)$ is not
reduced to a $T$-independent distinguish in the parameters
accounting for the $M_{\mbox{\small ZFC}}(T)$ and
$M_{\mbox{\small FC}}(T)$ dependences. One may expect that
observation of $\delta M$ here is related to the two-phase
coexistence. The $M_2$-measurements reveal the two-phase state
below $T^{\ast} \approx$ 143 K as the anomalous nonlinear
response becomes larger than the conventional one. Here we use
another way for disturbance of the system and analyze another
quantity, so that the possible inhomogeneous magnetic state may
be observed above $T^{\ast}$.  For instance, the increasing of
$\delta M$ on cooling below 220 K may reflect the differences in
the $T$-dependences of the volume fractions and/or mutual
arrangement of the phases. In addition, $T^{\ast}$ seems to be
also seen in $\delta M(T)$ data as a temperature where $\partial
\delta M(T)/\partial T$ has a maximum. At last, $\delta M(T)$
exhibits the maximum at $T_C$.  This corresponds to a conclusion
based on the $M_2$-data \cite{RyL} that the ordering of both
phases occurs simultaneously at $T_C$. These observations agree
with an assumption that formation of the two-phase state takes
place independently of the structural transition and does not
reduce to the coexistence of the $O^{\prime}$ and a small amount
of the initial $O$ phases.

Note some peculiarities of the $T$-hysteresis below $T_C$.
According to Fig.3a, difference between the FC and ZFC data
increases for $T <$ 90 K. This agrees with $T$-dependence of the
hysteresis loop found in the Re$M_2(H,T)$ measurements. The
modification seems to reflect change in a domain formation. Below
20 K the $M(T)$ exhibits the sharp reduction, indicating onset of
ordering of the Nd ions whose magnetization directed antiparallel
to that of the Mn moments.

\section{ESR}

Fig.4 shows the typical spectra for some temperatures obtained at
different $T$-scans. The similar spectra are also observed for
other temperatures above $T_C$. The spectra recorded at the
different treatments well coincide for temperatures above
approximately $T^{\ast}$, and are fitted by a Lorentzian line
shape
%- - - - - - - - - - - - - - - - - - - - - - - - - - - - - - - - - -
\begin{eqnarray}
\chi_{as}(\omega ,H) = \frac{1}{2}\chi
\biggl\{ \Bigl[\frac{\omega \Gamma}{(\omega-g\mu H)^2+\Gamma^2}-
\frac{\omega \Gamma}{(\omega+g\mu H)^2+\Gamma^2}\Bigr]
\mbox{cos}\varphi_s- \nonumber\\
\Bigl[\frac{\omega (\omega-g\mu H)}{(\omega-g\mu H)^2+\Gamma^2}-
\frac{\omega (\omega+g\mu H)}{(\omega+g\mu H)^2+\Gamma^2}\Bigr]
\mbox{sin}\varphi_s \biggr\}.
\end{eqnarray}

%-------------------------------------------------------------------

\noindent
Here $\chi_{as}$ is the antisymmetric component of the dynamic
susceptibility of the sample, $\chi$ is the static
susceptibility, $\omega$ is the frequency of $ac$-field, $\Gamma$
is the relaxation rate of a uniform magnetization, and
$\varphi_s$ is the phase of the signal. Since $\Gamma$ is a
rather large, the both circularly polarized components of $ac-$
field are taken into account in Eq.4. On this reason, a control
sample (a polycrystalline stable nitroxyl radical with $g$ =
2.055 and $\Gamma \approx $ 60 Oe) is used for tuning a phase of
$ac$ field which is the reference point for $\varphi_s$. The
fitting parameters in Eq.4 are $\Gamma$, $g$, $\varphi_s$ and the
amplitude of the signal $A_{as}(T) \propto \chi$.

We consider first a range $T^{\ast} \div 2T_C$ where the critical
behavior follows to that of a 3D isotropic ferromagnet
\cite{RyL}, and the structural refinement does not show any
structural transformation. Fig.5 presents the $\Gamma(T)$ and
$A_{as}(T)$ dependencies for the different $T$-scans which are
obtained between $T^{\ast}$ and 2$T_C$. It is seen that the
different treatments of the sample have a little effect on the
data. The $g$-factor (=2.00(1)) is independent of temperature
here.  $A_{as}(\tau)$ is expected to obey a scaling law for a 3D
isotropic ferromagnet, $A_{as}(\tau) \propto \tau^{-\gamma},
\gamma \approx$ 4/3, in a weak field regime at $\tau > \tau_H$
= $(g\mu H/kT_C)^{3/5} \approx 3.1\cdot 10^{-2}$ for resonance
field $H \approx$ 3 kOe. The inset in Fig.5b shows that
$A_{as}(\tau)$ curve follows this prediction down to
approximately $\tau \approx$ 0.11 and deviates from it at $\tau
<$ 0.11. This behavior is in a reasonable correspondence with
that of $\chi(\tau)$ found in the $M(H)$ (see above) and low
frequency measurements of the linear response on a weak $ac$
field \cite{RyL}.

Let us go to $\Gamma(\tau)$. In the weakly anisotropic cubic
ferromagnets, it is usually given by \cite{Hub,Mal,RLK}
%- - - - - - - - - - - - - - - - - - - - - - - - - - - - - - -
\begin{equation}
\Gamma(\tau) = \Gamma_c\tau^{-1} + \Gamma_{unc}\tau^{4/3}.
\end{equation}

%- - - - - - - - - - - - - - - - - - - - - - - - - - - - - - -

\noindent
Here $\Gamma_c$ is controlled by the dipolar forces, a single ion
anisotropy ($\propto S^{\alpha}S^{\beta}$) and anisotropy of the
exchange interaction (the two latter appear in the $Pbnm$
symmetry) \cite{HAC,DKS}. The term with $\Gamma_{unc}$ describes
the uncritical contribution including a spin-lattice relaxation
rate. The Dzyaloshinsky-Moria interaction also presents in this
space group \cite{HAC,DKS}. Following to an analysis given in
Ref. \cite{Laz}, one can show that this interaction gives
contribution in $\Gamma_{unc}$ only since it connects
fluctuations of the magnetization (critical) and staggered
magnetization (uncritical) whose characteristic energy $\sim
T_C$. The $\Gamma(\tau)$ exhibits a critical enhancement at
$\tau \rightarrow$ 0 below a minimal value at $\tau_m$. This
behavior is observed in the traditional cubic ferromagnets
\cite{BIL} and in the near to cubic manganites
La$_{1-x}$Ca$_x$MnO$_3 (x = 0.18 \div$ 0.22) with $\tau_m \sim$
0.25 \cite{SRG}. In our compound $\Gamma(\tau)$ reduces
monotonically at $\tau \rightarrow 0$. This dependence cannot
be described by the single term with $\Gamma_{unc}$ and suggests
the presence of a new relaxation mechanism. Since Nd$^{3+}$ ion
has a magnetic moment and this magnetic subsystem is not
presented in the indicated above compounds, a Mn-Nd coupling is
the most likely to be additional source of the relaxation. The
ground state of Nd$^{3+}$ ion (total moment $J$ = 9/2, $L$ = 6
and $S$ = 3/2) is split by a low symmetric (lower than cubic)
crystal field into the five doublets described by the effective
spins $S$ = 1/2. It is expected that a Mn-Nd exchange
interaction $\hat J_{12}$ is larger than that of the Nd-Nd
\cite{KSR,PMR}. We will show that the weak anisotropic coupling
$\hat J_{12}$ can give the needed contribution in
$\Gamma(\tau)$. The main effect of $\hat J_{12}$ can be
explained by taking into account the single ground doublet of
the Nd$^{3+}$ ion and neglecting the Nd-Nd interactions. We
consider first a result of a perturbation theory for $\Gamma$
with $H_{int} = \sum\nolimits_{i,k}\mbox{\bf S}_{1i}\hat
J_{12ik}\mbox{\bf S}_{2k}$, where {\bf S}$_{1i}$ and {\bf
S}$_{2i}$ are spins of the Mn and Nd ($S_2$ = 1/2) ions,
respectively. A relaxation rate $\Gamma_{MnNd}$ of
total spin of the system {\bf S = S$_1$ + S$_2$} is determined
by an anisotropy of $H_{int}$. To get $\Gamma_{MnNd}$, a pair
spin Green function of the operators $\partial
S^{\alpha}/\partial t$ is analyzed \cite{Mal}. This includes a
decoupling of a four-spin Green function, which appears as a
result of a commutating [$S^{\alpha},H_{int}$] into product of
the two-spin functions. In our case the latter are the Green
functions of the Mn ($G_1$) and Nd ($G_2$) subsystems, $G_1$
having the properties of a 3D isotropic ferromagnet and $G_2$
corresponding to the free Nd spins. The
$\Gamma_{MnNd}$ is also determined by an uniform
static Green function of the system $G^{(0)} = G_{1}^{(0)} +
G_{2}^{(0)}$ ($\Gamma_{MnNd} \propto$
$(G^{(0)})^{-1}$, $G^{(0)} \approx G_{1}^{(0)}$ at a small
$\tau$ because $G_{1}^{(0)} \sim T_{C}^{-1}\tau^{-4/3}$ and
$G_{2}^{(0)} \sim T^{-1}$). As a result, we find
$\Gamma_{MnNd}(\tau) \propto \tau^{1/3}$. This
answer differs from the critical term ($\propto \tau^{-1}$,
Eq.(2)) in a factor $\tau^{4/3}$ since the critical contribution
contains product of the two critical functions ($G_{1}^2$)
whereas $\Gamma_{MnNd} \propto G_{1}^{(0)}G_2^{(0)}$ that gives
$G_{2}^{(0)}/G_{1}^{(0)} \propto \tau^{4/3}$. The magnitude of
$\Gamma_{MnNd}$ is determined by anisotropic part
of the $\hat J_{12}$({\bf q}) at {\bf q} = 0, where $\hat
J_{12}$({\bf q}) is the Fourier transform of $\hat J_{12ik}$,
$\hat J_{12}(0)$ = $Z_{12}\hat J_{12} \equiv \hat  U_{12},
Z_{12} \approx$ 8 is the coordination number of the Mn-Nd
sublattice.  This part is $(U_{12}^{an})^{\alpha
\beta} = U_{12}^{\alpha \beta} - \delta_{\alpha \beta}\bar
U_{12}$, where $\bar U_{12} =$ Sp$\hat U_{12}/3$ is the mean
exchange coupling. Generally speaking, one can find the
different relaxation rates along the axes of the crystal.
However, the anisotropy of $\Gamma$ is not observed. It can be
due to a twinning of the crystal in the ($a,b$) plane and an
averaging of the relaxation rates by the circular polarized
components of $h(t)$. Therefore, we introduce an effective
exchange anisotropy $(U_{12}^{an})^{\alpha \beta}
\rightarrow U_{12}^{an}$ and write
%- - - - - - - - - - - - - - - - - - - - - - - - - - - - - - -
\begin{equation}
\Gamma_{MnNd} =
\Gamma_{MnNd}^{an}\tau^{1/3}, \quad
\Gamma_{MnNd}^{an} \sim (U_{12}^{an})^2/T_C.
\end{equation}

%- - - - - - - - - - - - - - - - - - - - - - - - - - - - - - -

\noindent
$\Gamma_{MnNd}^{an}$ reflects an
exchange narrowing that is due to the strong isotropic exchange
interaction in the Mn subsystem. This law is the consequence of
decay of the uniform magnetization into the critical and
uncritical modes which is enforced by the anisotropy, a
characteristic energy of the free Nd-spins being obviously equal
to zero. Note that the same contribution in $\Gamma$ appears
also as an uncritical (or rather intermediate) mode has an
uncritical amplitude ($\sim T_C$) and a characteristic energy
that is equal that of the critical mode. This occurs in some
cubic collinear ferrimagnets \cite{BeI}.

To determine a region of the validity for this result, the
equations of motion for the Green functions $\hat G_1$({\bf
q},$\omega$) and $\hat G_2$({\bf q},$\omega$) at {\bf q} = 0
should be considered. The $\hat U_{12}$ introduces a frequency
scale $\Omega_2(\tau)$ in $\hat G_2(\omega)$ that gives the main
restriction. The $\hat U_{12}$ is assumed to be rather
anisotropic $\mid (U_{12}^{an})^{\alpha \beta}\mid \;
\sim \; \mid\bar U_{12}\mid$. This is not unusual situation for
the exchange coupling between the $d$ and $f$ ions in the
perovskite lattice and an estimation given below supports this
assumption. Therefore, one can use the isotropic component
$\bar U_{12}$ to get $\Omega_2(\tau)$. The traditional mode
coupling theory \cite{Mal} yields $\Omega_2(\tau) \sim (\bar
U_{12}^2/T_C)\tau^{-1}$. Result (6) is valid if $\Omega(\tau) =
T_C\tau^{5/3} > \Omega_2(\tau)$, i.e. $\tau > \tau_{an} \sim
(\mid \bar U_{12}\mid /T_C)^{3/4}$, $\mid \bar U_{12}\mid \quad
\sim \quad \mid (U_{12}^{an})^{\alpha \beta}\mid$, where
$\Omega(\tau)$ is the characteristic energy of the critical
fluctuations for a 3D isotropic ferromagnet. At $\tau <
\tau_{an}$, a new dynamic regime originates which
corresponds to a system with completely destroyed conservation
law of the {\bf S}. This regime develops when the static
behavior weakly changes. Employing the usual two sublattice
expressions for $G^{(0)}$ and $\bar U_{12}\delta_{\alpha \beta}$
for the estimations, we find for a static anisotropic border
$\tau_{an}^{st} \sim \tau_{an}^2$. Note that the small
anisotropic spin interactions in the Nd subsystem
$U_{NdNd}^{an}$ (for instance, the dipolar forces) give an
additional contribution in $\Gamma_{unc}$. They do not effect
$\tau_{an}$ if $\Omega(\tau_{an}) > \mid (U_{NdNd}^{an})^{\alpha
\beta}\mid$.

We have discussed above dynamics of the {\bf S} which coincides
with that of the total magnetization {\bf M} when the $g$-factors
of the Mn ($g_1$) and Nd ($g_2$) ions are isotropic and equal to
each other. The anisotropy of $\hat g_1$ can be neglected whereas
that of $\hat g_2$ may be substantial. Consider first a case of
isotropic $g_1 \neq g_2$. When $\hat J_{12}$ = 0, the spectrum
reveals the two resonance at $\omega_{1,2} = g_{1,2}\mu H$. With
increasing of isotropic part of the Mn-Nd exchange, the lines
begin to overlap. In a regime of a strong exchange narrowing
($\mid \bar U_{12} \gg g_{1,2}\mu H$), the spectrum is dominated
by a single resonance determined by an effective $g-$factor
$g_{eff}$ \cite{DHu}. Contributions of the $g_1$ and $g_2$ in
$g_{eff}$ depend on the relationships between $G_{1}^{(0)},
G_{2}^{(0)}$ and $\mid \bar U_{12}\mid $ as well as $S_1$ and
$S_2$. When $G_{1}^{(0)}$ dominates ($G_{1}^{(0)} \gg
G_{2}^{(0)}, g_{eff} \approx g_1$ if $\mid \bar U_{12}\mid
G_{2}^{(0)} \leq$ 1 and $S_1 \geq S_2$. In virtue of the same
reason, the anisotropy of $\hat g_{eff}$ related to that of
$\hat g_{2}$ and $\hat U_{12}$ is also to be small. The
expression for $\Gamma_{MnNd}$ (6) corresponds to
the damping of this resonance. Note, that in our case the
admixture of $g_2$ to $g_1$ is very small: $g_{eff} = g_1 +
Cg_2$ \cite{DHu}, where $C \sim [S_2(S_2 + 1)/S_1(S_1 +
1)]^2(G_{2}^{(0)}/G_{1}^{(0)}) \approx 4\cdot 10^{-4}$ even at
$\tau$ = 1 because $G_{1}^{(0)} = C_{\chi}S_1(S_1 + 1)/3T_C$,
$C_{\chi} \approx$ 2.95, $S_1 \approx$ 1.88 and $G_{2}^{(0)} =$
$C_{Nd}S_2(S_2 + 1)/6T_C, C_{Nd}$ = 0.75 is the concentration
of Nd atoms and $\mid\bar U_{12}\mid G_{1}^{(0)} \ll$ 1 for
estimated below $\mid\bar U_{12}\mid \quad \sim$ 5 K.

The same factor $C$ accounts for the $T$-dependent corrections to
$g_1$ related to the Mn-Nd coupling. The closeness $g_1$ =
2.00(1) to $g_1 \approx$ 2 for La$_{1-x}$Ca$_x$MnO$_3$
\cite{SRG} and its independence from $T$ agree well with the
smallness of $C$.

Going to the data, we present $\Gamma$ by
%- - - - - - - - - - - - - - - - - - - - - - - - - - - - - - -
\begin{equation}
\Gamma(\tau) = \Gamma_c\tau^{-1} +
\Gamma_{MnNd}^{an}\tau^{1/3} + \Gamma_{unc}\tau^{4/3}.
\end{equation}

%- - - - - - - - - - - - - - - - - - - - - - - - - - - - - - -

\noindent
Inset in Fig.5a shows the fit of the $\Gamma(\tau)$ for 1 $> \tau
>$ 0.13. The fitting parameters are $\Gamma_c$ = 9(2) Oe,
$\Gamma_{MnNd}^{an}$ = 590(20) Oe and $\Gamma_{unc}$ = 505(24)
Oe. The $\Gamma_c$ is controlled by the dipolar forces since the
corresponding contribution is estimated as $\Gamma_d \sim 4\pi
(g\mu)^2/V_0T_C \approx$ 16 Oe for $V_0 \approx (3.9)^3$ \AA$^3$
\; \cite{BIL}. It is remarkable that for $\Gamma_{MnNd}^{an}$ =
0 we obtain $\Gamma \approx$ 115 Oe at $\tau \approx$ 0.25. This
value is very close to $\Gamma = 120 \div$ 130 Oe which was
found in La$_{1-x}$Ca$_x$MnO$_3$ ($x = 0.18 \div$ 0.22 and $T_C$
= 180 K $\div$ 188 K) at $\tau_m \sim$ 0.25 \cite{SRG}. This
observation strongly supports the assumption on the leading role
that plays the Mn-Nd coupling in the suppression of critical
enhancement of the $\Gamma(\tau)$.  From Eq.(6), we find $\mid
U_{12}^{an}\mid \; \sim$ 3.2 K. The strength of $\hat
U_{12}$ is not completely known. We will try to estimate $\bar
U_{12}$ exploiting $T$-dependence of the small Nd magnetic
moment (0.45 $\mu$ at 1.5 K) which was measured by neutron
diffraction in Nd$_{0.7}$Ba$_{0.3}$MnO$_3$ for 20 K $> T
\geq$ 1.5 K \cite{FSM}. This dependence suggests that Nd
magnetization is induced by the ordered Mn subsystem, i.e. it is
due to the splitting of the Nd ground Kramers doublet by the
mean field of the Mn spins.  Assuming that $\bar U_{12}$
dominates and $g_2 \sim$ 2, we find $\mid\bar U_{12}\mid \;
\sim$ 5 K of the same order as $\mid\bar U_{12}^{an}\mid$.

The border of the anisotropic dynamical crossover is
$\tau_{an} \sim (\mid\bar U_{12}^{an}\mid/T_C)^{3/4}
\approx 6.3\cdot 10^{-2}$. The weak anisotropic spin
interactions in the Nd subsystem do not affect
$\tau_{an}$ if $T_C\tau_{an}^{5/3} \approx$ 1.3 K $> \;
\mid(U_{MnNd}^{an})^{\alpha \beta}\mid$. This condition
satisfies for the Nd-Nd dipolar interaction whose magnitude does
not exceed that of the Mn-Mn ($\sim$ 0.5 K). The exchange
coupling of the Nd spins seems to be an order of 1 K or less
\cite{GGH}. The $\tau_{an}$ is close to the
minimal $\tau \approx$ 0.13 which is available for the precise
analysis of the $\Gamma(\tau)$. Since $\Gamma(\tau)$ cannot
exhibit a critical enhancement in the anisotropic regime, this
effect is completely destroyed in our compound that is confirmed
by estimation of the line width at $\tau <$ 0.13.

The anisotropic static crossover occurs at a very small
$\tau_{an}^{st} \sim \tau_{an}^2 \approx 4\cdot 10^{-3}$
($T_{an}^{st} = T_C$ + 0.5 K). It is not surprising, therefore,
that $\chi(\tau)$ dependence of our compound and the static
critical properties of Nd$_{0.6}$Pb$_{0.4}$MnO$_3$ ($T_C
\approx$ 156 K) \cite{SRN} correspond to those of a 3D isotropic
ferromagnet.

We have characterized above the Nd$^{3+}$ ion by its single
ground doublet. Other states are ineffective if a distance
between the ground doublet and a next one is larger than 2$T_C$.
It can be a realistic situation, since a cubic component, which
dominates in the crystal field of the weakly distorted cubic
crystal, can produce a large splitting between the ground doublet
and the two upper quartets. In a comparably strongly distorted
NdCuO$_4$, for example, this splitting is 15 meV whereas the
total splitting is 109 meV \cite{LoS}. If the next several
states are relevant they can be simply taken into account since
a distance between the neighboring doublets is definitely larger
than Mn-Nd exchange. In this case these states can be treated
separately by the same manner as the ground state so that they
produce the obvious additive effect.

We consider now the ESR data below $T^{\ast}$ where a pronounced
deviation of the signal from Lorenzian starts to be observed. A
reason is that Im4$\pi \chi_{xx}(\omega, H) \gg$ 1 at $\omega
\approx g\mu H$ as 4$\pi \chi \sim$ 1 (4$\pi \chi \approx$ 0.6 at
$T^{\ast}$). This leads to a complex nonuniform distribution of
the $ac-$field in the sample because a wave vector $k$ in a
medium $k \propto \{\mu_0 = 1 + 4\pi
\chi_{xx}(\omega, H)\}^{1/2}$ for a wave spreading along {\bf H},
where, for simplicity, we neglected the $\chi_{as}(\omega, H)$.
As a result, phase $\varphi_s$ of the signal begins to exhibit a
strong $H$-dependence that cannot be correctly determined for our
sample in the resonator. Therefore below $T^{\ast}$, we present
directly the signals for the several types of the $T$-scans. The
pronounced $T$-hysteresis observed in all the measurements
(Fig.4) is the clear indication of the anomalous behavior that
develops below $T^{\ast}$. Since a precise quantitative analysis
of the signal is not feasible here, we cannot determine whether
this hysteresis is caused by that of $M$ or it is related also to
hysteresis of the $\Gamma$ and $g-$factor. A rough estimation
gives approximately $T$-independent values for $\Gamma$ and
$g-$factor below $T^{\ast}$ down to $T_C$. The same reason does
not allow us to elucidate a possible two-phase character of the
signal. At the same time, above $T^{\ast}$ phase $\varphi_s$
reflects mainly phase shift of the $h(t)$ in the sample which is
related to a comparably large $\varepsilon \sim$ 10 and
conductivity $\sigma$. This effect, being independent of $H$, can
be taken into account correctly by Eq.(4). For the largest
conductivity ($\sigma \approx 1.7 (\Omega$cm)$^{-1}$) at $T =
2T_C \approx$ 260 K, we get $l_{\perp}\mid k\mid \; \sim$ 0.25
for $\mu_0 \approx$ 1, were $l_{\perp} \approx$ 0.1 mm is the
thickness of the sample. Since $\sigma$ decreases sharply on
cooling ($\sigma \sim 0.3\cdot 10^{-2}$ ($\Omega$cm)$^{-1}$ at
$T^{\ast}$), the $h$-distribution in the sample is closed to
uniform between $T^{\ast}$ and $2T_C$.

\section{Discussion and conclusion}

The $T$-hysteresis observed in the magnetization and ESR
measurements above $T_C$ evidences at least a weak first order
transition. According to the $M_2$-data, a reason of changing
regime of the transition at $T^{\ast}$ is formation of the new
phase with the anomalous strong nonlinear magnetic behavior in
the weak fields which is stronger than that of a 3D isotropic
ferromagnet near $T_C$. This phase coexists with the normal one
at least from $T^{\ast}$ down to $T_C$. This PS magnetic state
cannot be trivially related to a two-phase structure because,
according to the neutron diffraction, the sample remains the
structural uniform compound below $T^{\ast}$. Note that, first,
the anomalous phase coexists with the conventional one possessing
the different orbital ordering for the $x$ = 0.23 and $x$
= 0.25 compounds. Second, it arises in the manganites that are
closed to border of the I-M transition ($x \approx$ 0.3), where,
according to Ref.\cite{FSM} and our preliminarily data, the
orbital ordering disappears. These observations suggest that the
anomalous behavior can be attributed to an orbital disordered
metallic phase. The lack of observability of the anomalous phase
in the structural measurements means that this phase is
structurally closed to the conventional one and/or its volume is
small. Fragmentation of the sample in the PS state may be such
that the anomalous phase does not form a percolative conductive
cluster, and the system remains an insulator. The strong
$H$-nonlinear behavior of this phase is expected to be a combined
effect of its own properties and a magnetic coupling of its
fragments through the normal phase with the developed
ferromagnetic correlations. It is very important to stress a
difference between this transition and a first order one whose
character is determined on the base of $M(H)$ data. In the
latter, type of the transition (first/second) is characterized by
sign of coefficient $b$ in an expansion of $H/M$ in $M^2$ above
$T_C$: $H/M = a + bM^2 + \ldots$ ($b <$ 0 (first order), $b >$ 0
(second order)). Such a criterion suggests that the magnetization
is the single critical degree of freedom accounting for the
transition. This peculiarity is indeed observed in the $M(H)$
dependence of traditional ferromagnet MnAs \cite{BeR} and a
number of the manganites \cite{MRR,HKH}. Besides the
conventional $M(H)$ measurements, sign of $b$ can be found from
the $M_2$-data because Re$M_2(\omega, H) \propto
\partial^2M/\partial H^2 = - \; 6(b/a^4)H$ for $H \rightarrow$ 0
and a small $\omega$ when Re$M_2$ dominates in the response
\cite{RLL}. Our results on the $x$ = 0.23 and $x$ = 0.25
compounds obtained by this method indicate clearly that $b >$ 0
for both coexisting phases \cite{RLL,RyL}. This implies that $M$
is not the single degree of freedom involved in the transition,
and one should consider the charge, orbital and JT-phonon
variables (or some of them) as the critical ones as well. It is
not the unexpected conclusion for the manganites that are
characterized by close interconnection of the magnetization and
these degrees of freedom. In our system, the charge and orbital
degrees of freedom together with $M$ are the most likely to be
the critical variables. The JT-subsystem seems to be less
important because the JT distortions are small.

\section* {\small ACKNOWLEDGMENTS}

We thank V.I. Voronin for help in magnetization measurements. We
are also grateful to V.P. Khavronin for helpful discussions. This
work was supported from the joint Russian-Belorussian Foundation
for Basic Research, Grants No.04-02-81051 RFBR-Bel2004\_a,
F04R-076 and partly from the State Program of neutron
investigation (NI) as well as Russian Foundation for Basic
Research, Grant No. 02-02-16981.

%====================================================================

%\end{multicols}
%\widetext

\newpage
{\bf Figure captions}

\noindent
%\begin{figure}
%\centerline
%\hfill{\epsfxsize=8cm \epsfysize=6cm
%\epsfbox{fig1.eps}}
%\caption{
Fig.~1. The neutron diffraction pattern, calculated profile and
residual curve for Nd$_{0.75}$Ba$_{0.25}$MnO$_3$ at room
temperature. The insert shows the wide and weak monoclinic peaks
at the small angles.
%\label{figurename}}
%\end{figure}

\noindent
%\begin{figure}
%\centerline
%\hfill{\epsfxsize=8cm \epsfysize=6.5cm
%\epsfbox{fig2.eps}}
%\caption{
Fig.~2. Temperature dependences of the structural parameters $a,
b, c$ (panel (a)); volume of the unit cell $V$ and magnetic
moment of the sample per Mn ion (panel (b)); the Mn-O bond
lengths (panel (c)); and the Mn-O-Mn angles (panel (d)).
%\label{figurename}}
%\end{figure}

\noindent
%\begin{figure}
%\centerline
%\hfill{\epsfxsize=8cm \epsfysize=6.5cm
%\epsfbox{fig3.eps}}
%\caption{
Fig.~3. Temperature dependence of the magnetization for the
cooling and heating regimes under $H$ = 1 kOe. Panel (a) shows
these dependencies in the full temperature range 6 $\div$ 300 K.
Insert (1) displays fit of the $M^{-1}(T)$ measured in the ZFC
regime to the function $\tau^{\gamma}/CH + 4\pi N/H$. The fitting
parameters are found to be $1/CH =$ 3.98(3) g/emu and $\gamma$ =
1.39(1) at $N$ = 1/3. Insert (2) represents fit of the ZFC
$M(\tau)/H_{int}$ to expressions (2), (3) (see the text) in
$T$-range $T_C \div$ 261 K. Panel (b) shows the relative
difference $\delta M$ versus $T$ in the $T$-range 80 $\div$ 300
K. Insert in (b) displays the $M(T)$ curves registered in the ZFC
and FC regimes in $T$-range 140 $\div$ 180 K.
%\label{figurename}}
%\end{figure}

\noindent
%\begin{figure}
%\centerline
%\hfill{\epsfxsize=8cm \epsfysize=6.5cm
%\epsfbox{fig4.eps}}
%\caption{
Fig.~4. The ESR spectra for the different $T$-scans at some close
temperatures: cooling (solid line, panels (a-d)), heating after
fast cooling (dashed line, panels (a-c)), cooling under $H$
= 4 kOe (dotted line, panels (a) and (d)). Panels (a) and (c)
represent also fit of the spectra recorded on cooling to a
Lorenzian (dash-dotted line).
%\label{figurename}}
%\end{figure}

\noindent
%\begin{figure}
%\centerline
%\hfill{\epsfxsize=8cm \epsfysize=6.5cm
%\epsfbox{fig5.eps}}
%\caption{
Fig.~5. Temperature dependencies of parameters of the ESR spectra
for different $T$-treatments of the sample: cooling (full
symbols) and heating (open symbols) under $H$ = 0 (squares) and
$H$ = 4 kOe (triangles); heating after fast cooling (stars).
Panel (a) displays the spin relaxation rate $\Gamma$ versus
temperature. Insert shows $\Gamma(\tau)$ dependence on cooling,
and its fit described in the text. Panel (b) represents
amplitudes $A_{as}(T)$ of the spectra versus $T$. Insert shows
fit of $A_{as}(\tau) \propto \chi(\tau)$ to the scaling law.
%\label{figurename}}
%\end{figure}

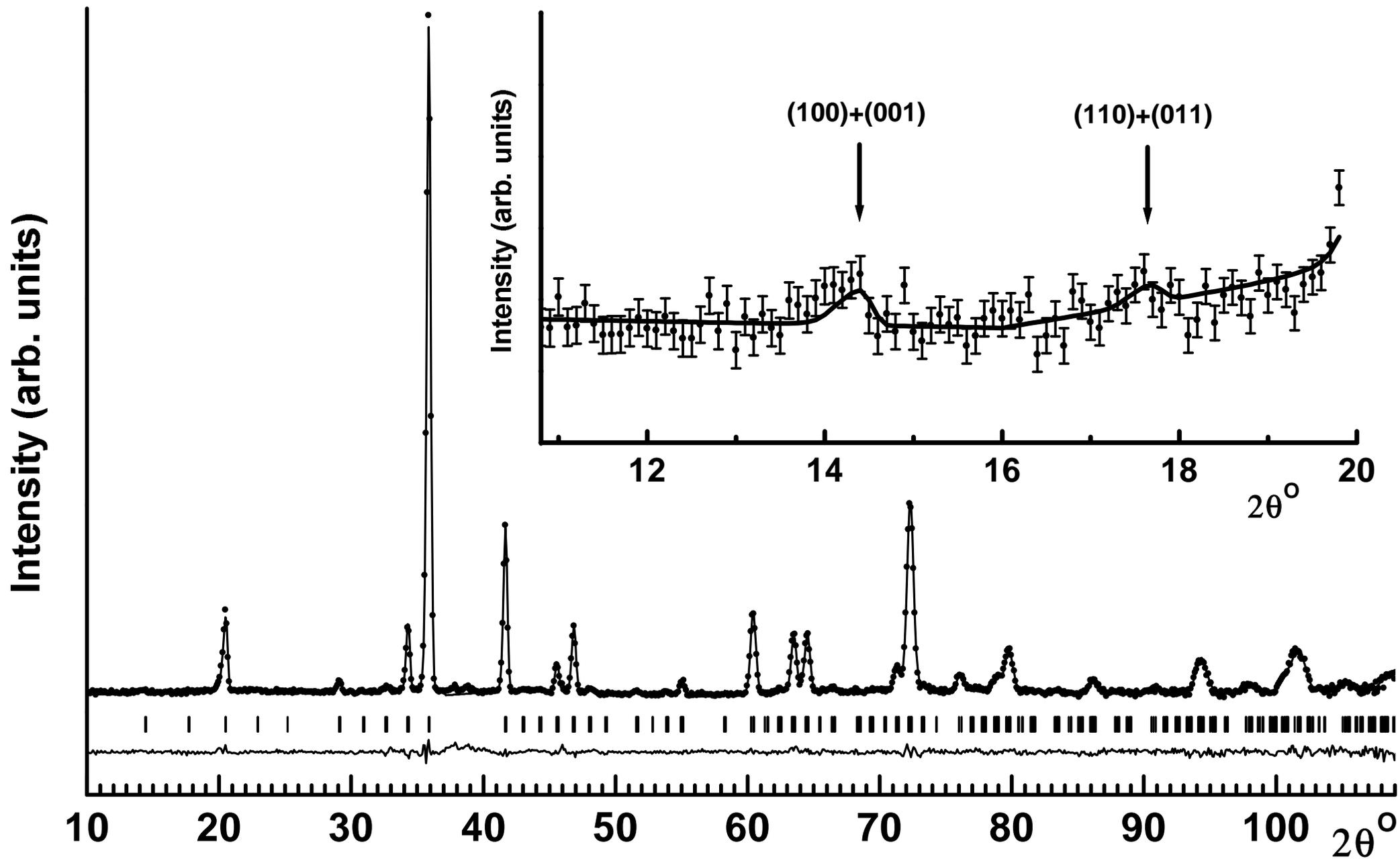
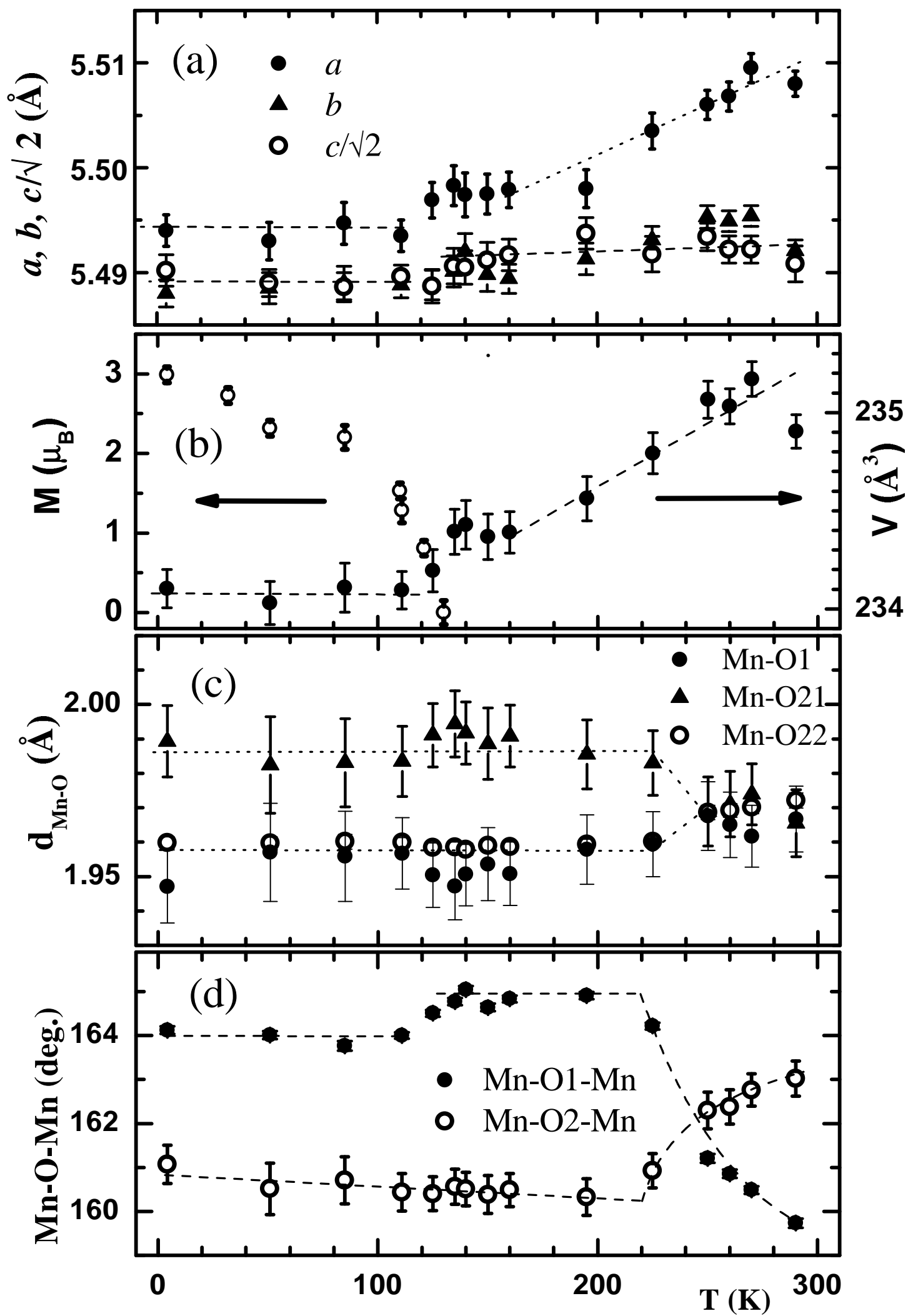
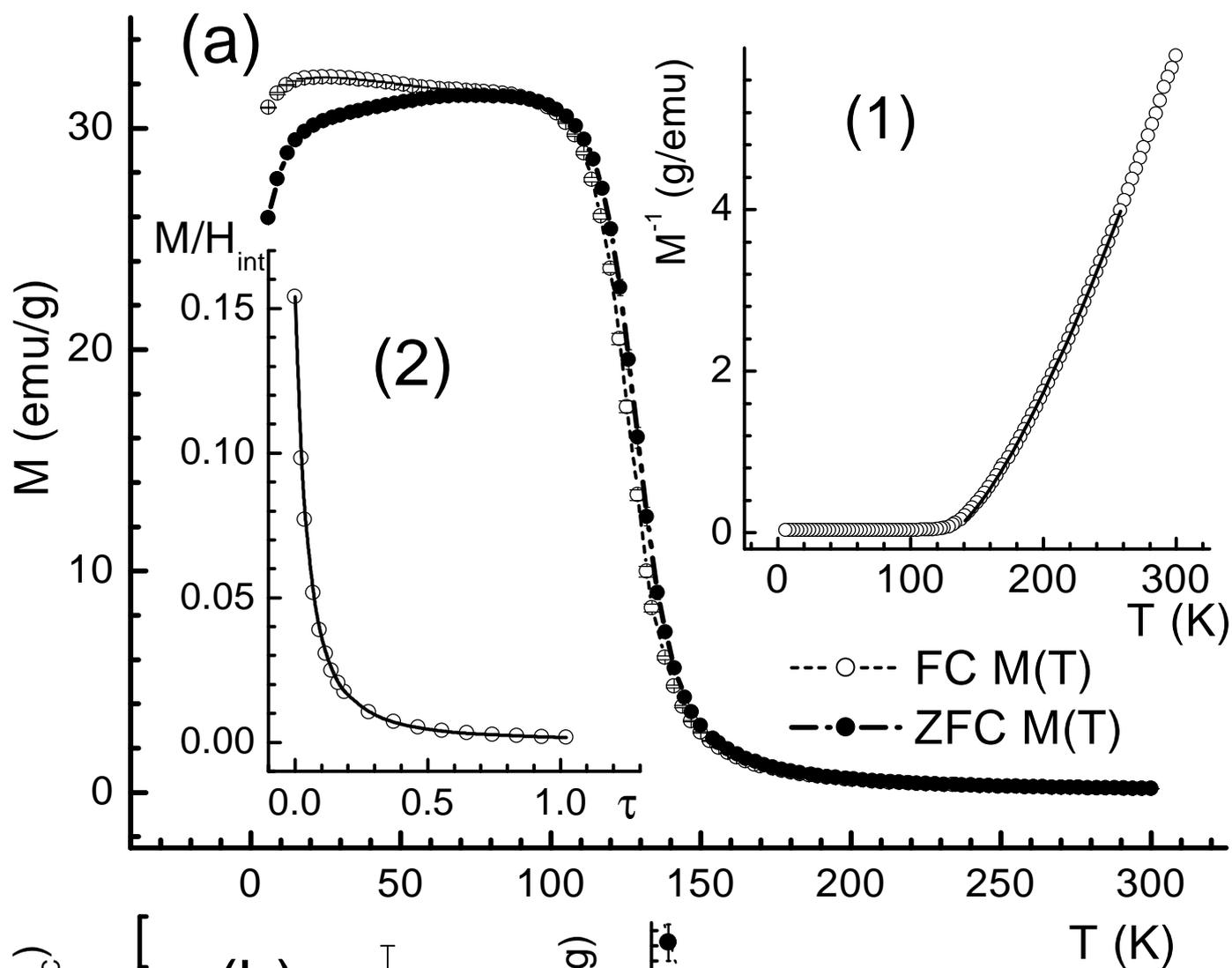
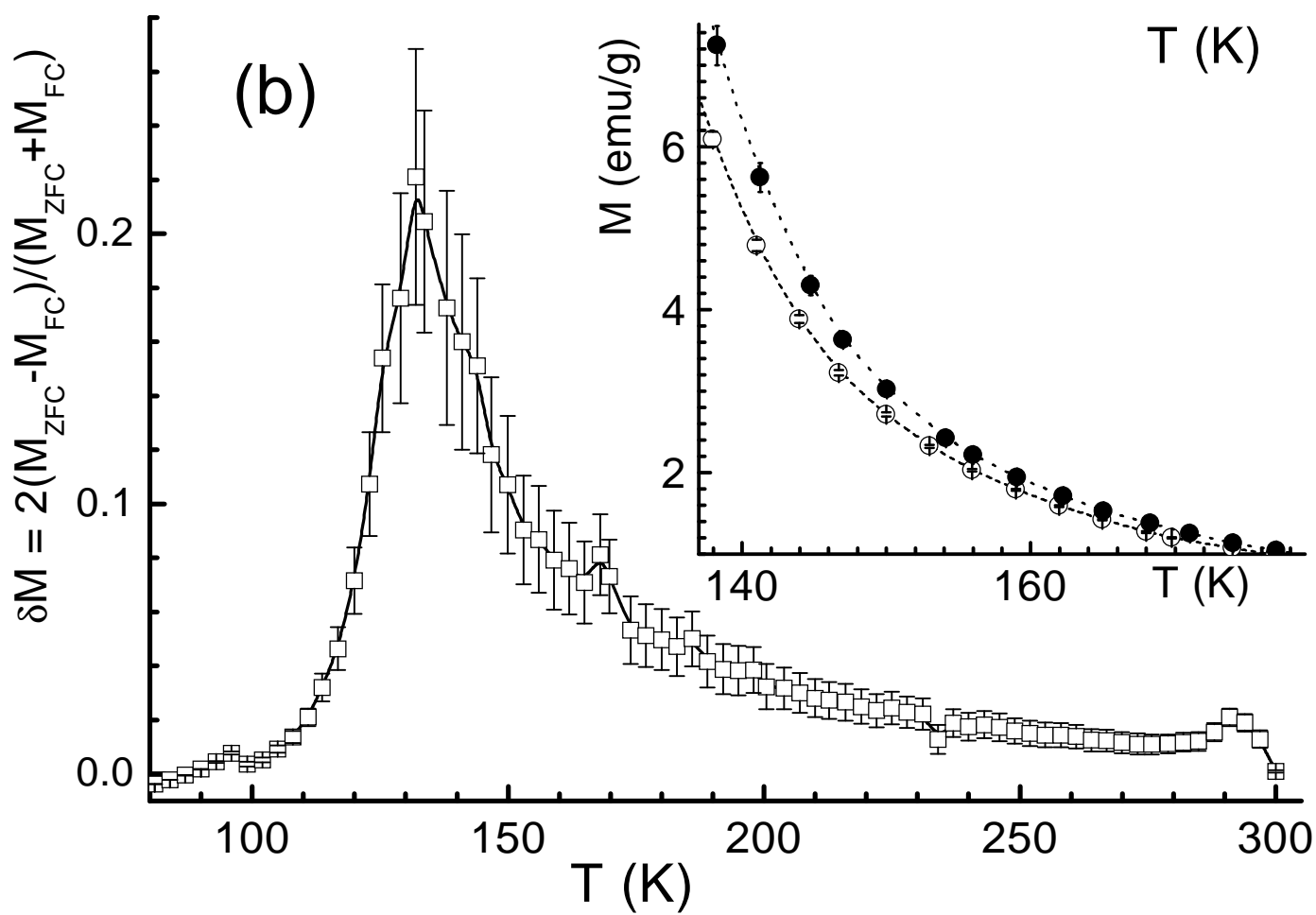
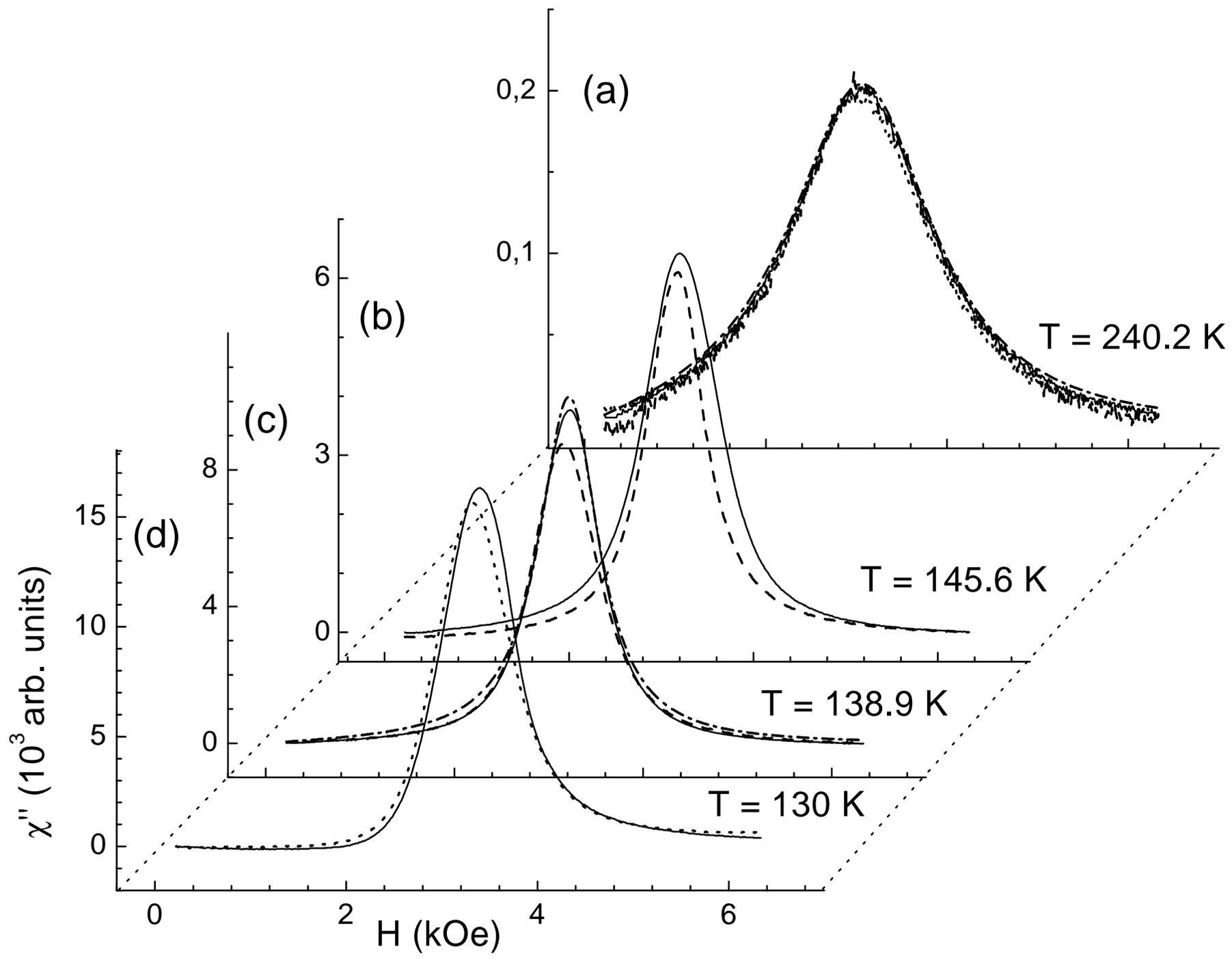
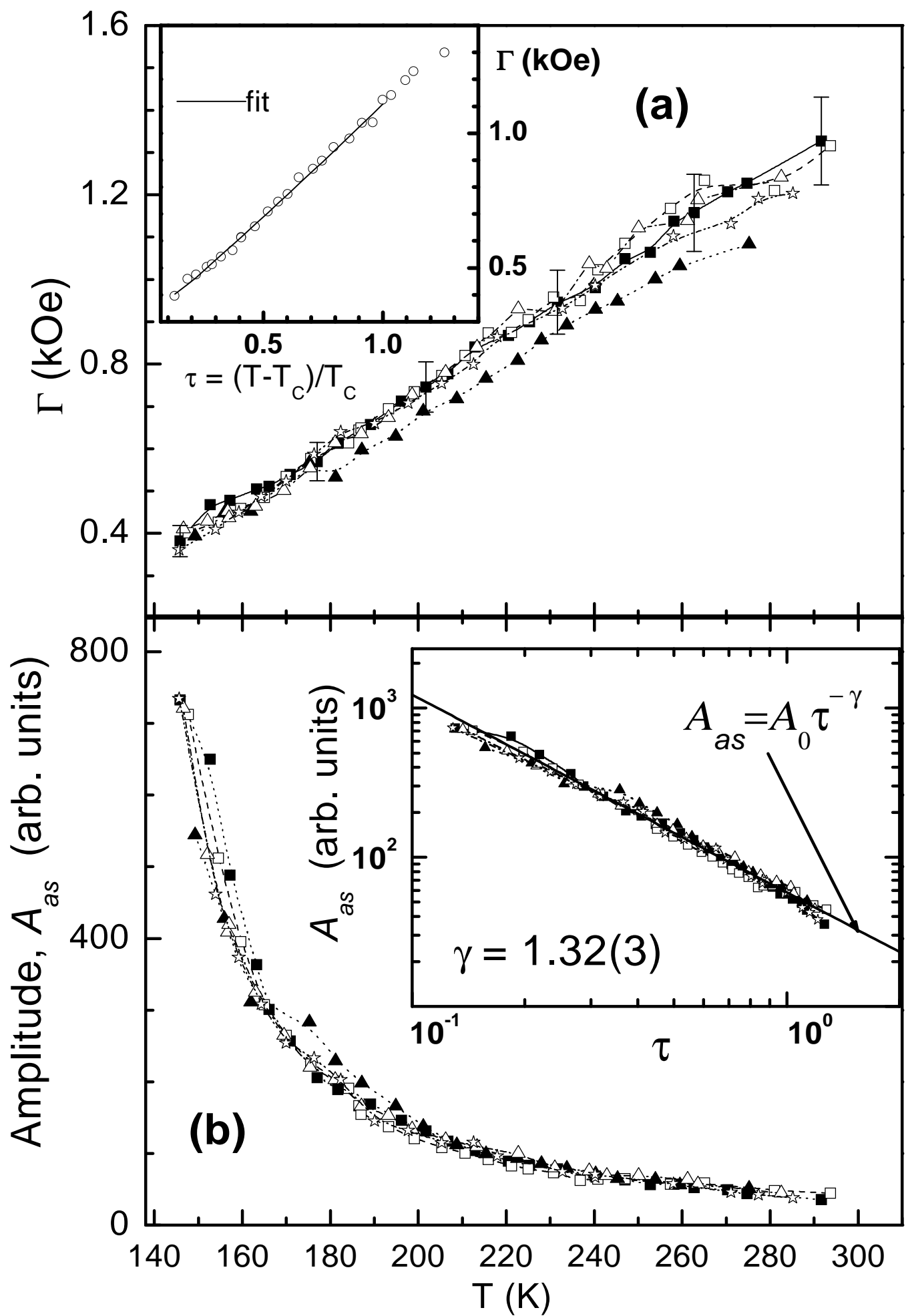
\end{document}